\documentclass{article}
\usepackage{spconf,amsmath,graphicx}
\usepackage{amssymb,url,times}
\usepackage{color}
\usepackage{enumitem}
\usepackage{multirow}
\usepackage{balance}
\usepackage{cite}
\usepackage{booktabs}
\usepackage[table,xcdraw]{xcolor}
\usepackage[caption=false]{subfig}
\usepackage{siunitx} %
\usepackage{adjustbox}
\usepackage{etoolbox}%
\robustify\bfseries
\makeatletter
\AtBeginDocument{\patchcmd{\sf@subfloat}%
  {\maincaptiontopfalse}%
  {\maincaptiontoptrue}%
  {}{}}%
\makeatother

\usepackage{pifont}%
\newcommand{\batsu}{{\footnotesize \ding{53}}}%

\title{WHAMR!: Noisy and Reverberant Single-Channel Speech Separation}
\name{Matthew Maciejewski$^{1,2}$,
      Gordon Wichern$^{1}$,
      Emmett McQuinn$^3$, 
      Jonathan Le Roux,$^{1}$\thanks{This work was performed while M.~Maciejewski was an intern at MERL.}}
\address{$^1$Mitsubishi Electric Research Laboratories (MERL), Cambridge, MA, USA\\ $^{2}$Johns Hopkins University, Baltimore, MD, USA \quad
$^3$Whisper.ai, San Francisco, CA, USA\\
         {\small\texttt{mmaciej2@jhu.edu, \{wichern, leroux\}@merl.com, emmett@whisper.ai}}
}
\begin{document}
\ninept
\maketitle
\setlength{\abovedisplayskip}{5pt}
\setlength{\belowdisplayskip}{5pt}
\begin{abstract}
While significant advances have been made with respect to the separation of overlapping speech signals, studies have been largely constrained to mixtures of clean, near anechoic speech, not representative of many real-world scenarios. Although the WHAM! dataset introduced noise to the ubiquitous wsj0-2mix dataset, it did not include reverberation, which is generally present in indoor recordings outside of recording studios. The spectral smearing caused by reverberation can result in significant performance degradation for standard deep learning-based speech separation systems, which rely on spectral structure and the sparsity of speech signals to tease apart sources. To address this, we introduce WHAMR!, an augmented version of WHAM! with synthetic reverberated sources, and provide a thorough baseline analysis of current techniques as well as novel cascaded architectures on the newly introduced conditions.
\end{abstract}
\section{Introduction}
\label{sec:intro}
In recordings produced in natural settings with multiple speakers present, it often occurs that more than one person will speak at the same time.
The resulting overlapped speech can cause a severe degradation in the performance of speech processing technologies designed for only a single speech signal, such as automatic speech recognition and speaker identification. Moreover, overlapped speech can be difficult to understand for human listeners as well. Speech separation systems aim to solve this problem by producing multiple waveforms, each estimating the clean speech of a single speaker, from recordings of overlapped speech.

Great advancements have been made in recent years on solving the speech separation problem through deep learning-based techniques~\cite{Hershey2016,Isik2016,Kolbaek2017,Wang2018ICASSP04Alternative,luo2019convTasNet,shi2019furcanext}. However, the overwhelming majority of research conducted thus far has used the wsj0-2mix dataset~\cite{Hershey2016}, which consists of synthetically-mixed studio recordings of read utterances from the WSJ0 corpus~\cite{garofolo1993csr} and is not representative of many real-world scenarios in which overlapped speech may be present~\cite{bengio2005machine}. In many cases where multiple people are speaking at the same time, they are not speaking directly into the microphone, and are instead captured by a microphone placed at some distance away in the room, as in meetings or in home settings. In these far-field conditions, the distance from the source to the microphone can lead to a relative increase in noise compared to the speech and to increased reverberation \cite{gannot2017perspective}, neither of which are present in the most common deep learning-based speech separation evaluations. %
The addition of noise not only masks the speech signal but also corrupts phase information, while reverberation causes spectral smearing of the source. These phenomena could be challenging for separation systems which rely on the spectral structure of speech in the time-frequency domain~\cite{Vincent2018textbook}.  The introduction of the WHAM! dataset~\cite{wichern2019wham}, consisting of two speaker mixtures from the wsj0-2mix dataset together with real ambient noise, was a first step in the direction of more realism. It did not however consider reverberation or more generally spatialization of the speech signals, despite the noise samples being recorded in stereo.

To aid in the development and evaluation of speech separation systems in even more realistic conditions, we introduce the WHAMR! dataset that adds reverberation to WHAM!'s noise augmentation of wsj0-2mix.  We have generated realistic room parameters which are used to generate room impulse responses that can produce reverberant audio waveforms for each source in a manner similar to the multi-channel version of wsj0-2mix introduced in \cite{Wang2018ICASSP04MultiChannel}, but with the microphone geometry constrained by the binaural recording setup used to collect the WHAM! noise corpus.  %
Although some noisy and reverberant speech separation datasets were introduced in~\cite{maciejewski2019data}, they are constructed using actual recordings of noisy and reverberant speech. As such, they lack ground truth for clean and anechoic speech. WHAMR! provides a contrasting and complementary data paradigm; similarly to other WSJ0-based speech separation datasets, WHAMR! is constructed synthetically, with artificially-mixed speech plus noise and artificial reverberation.  This synthetic construction provides the ground truth of all speech signals with and without reverberation, which is necessary to effectively train and evaluate deep learning-based systems.

In this paper, we investigate the performance of various systems for clean, noisy, reverberant, and noisy plus reverberant separation as well as enhancement (denoising and dereverberation) tasks based on the WHAMR! dataset, establishing strong baselines and proposing new cascaded combination systems that can be trained end-to-end. %

\section{WHAMR! Dataset}
\label{sec:data}

The WHAMR! dataset\footnote{Available at: \url{http://wham.whisper.ai}} is an extension of the WHAM! dataset \cite{wichern2019wham}, which is a noise-augmented version of the wsj0-2mix dataset \cite{Hershey2016}. The wsj0-2mix dataset consists of mixtures of utterances from the WSJ0 corpus, combined with random gain between 0 and 5 dB to create overlapping speech. There are four configurations: a \textit{min} condition where the mixture is trimmed to the length of the shorter utterance and the corresponding non-trimmed \textit{max} condition, both available at 8~kHz and 16~kHz sampling rate. The mixtures are partitioned into training, validation, and test sets of 20,000, 5,000, and 3,000 mixtures respectively. %
In the WHAM! dataset, each speech mixture from the wsj0-2mix corpus was associated to a randomly sampled excerpt from noises recorded with binaural microphones in various urban environments throughout the San Francisco Bay Area, and mixed such that the louder speaker was at a randomly selected SNR between $-6$ and $+3$~dB relative to the noise~\cite{wichern2019wham}.

\begin{table}[tbp]
\centering
\vspace{-.1cm}
\caption{Room impulse response parameter sampling distributions. Units for all parameters are meters with the exception of reverberation time~($T_{60}$) which is in seconds and angles in radians.}
\vspace{0.05cm}
\label{table:rir_params}
\begin{adjustbox}{max width=\columnwidth}

\subfloat{
\begin{tabular}{@{}cc|c@{}}
\toprule
\multirow{3}{*}{\textbf{Room}} & L & $\mathcal{U}(5, 10)$ \\
 & W & $\mathcal{U}(5, 10)$ \\
 & H & $\mathcal{U}(3, 4)$ \\ \midrule

\multirow{3}{*}{$\mathbf{T_{60}}$} & high & $\mathcal{U}(0.4, 1.0)$ \\
 & med. & $\mathcal{U}(0.2, 0.6)$ \\
 & low & $\mathcal{U}(0.1, 0.3)$ \\ \bottomrule
\end{tabular}
}
\quad
\subfloat{
\begin{tabular}{@{}cc|c@{}}
\toprule
\multirow{3}{*}{\begin{tabular}[c]{@{}c@{}}\textbf{Mic.}\\ \textbf{Center}\end{tabular}} & L & $\frac{L_\text{Room}}{2}+\mathcal{U}(-0.2, 0.2)$ \\[1pt]
 & W & $\frac{W_\text{Room}}{2}+\mathcal{U}(-0.2, 0.2)$ \\[1pt]
 & H & $\mathcal{U}(0.9, 1.8)$ \\ \midrule
\multirow{2}{*}{\begin{tabular}[c]{@{}c@{}}\textbf{Mic.}\\ \textbf{Array}\end{tabular}} & sep. & noise mic. separation \\
 & $\theta$ & $\mathcal{U}(0, 2\pi)$ \\ \midrule
\multirow{3}{*}{\textbf{Sources}} & H & $\mathcal{U}(0.9, 1.8)$ \\
 & dist. & $\mathcal{U}(0.66, 2)$ \\
 & $\theta$ & $\mathcal{U}(0, 2\pi)$ \\ \bottomrule
\end{tabular}
} \end{adjustbox}
\vspace{-0.6cm}
\end{table}

WHAMR! extends WHAM! by introducing reverberation to the speech sources in addition to the existing noise. Room impulse responses were generated and convolved  using pyroomacoustics~\cite{scheibler2018pyroomacoustics} according to the random room configurations shown in Table~\ref{table:rir_params}. Reverberation times were chosen to approximate domestic and classroom environments~\cite{gannot2017perspective} (as we expect these to be similar to the restaurants and coffee shops where the WHAM! noise was collected), and further classified as high, medium, and low reverberation based on a qualitative assessment of the mixture's noise recording.

We created spatialized versions---\textit{anechoic} and \textit{reverberant}---of all components of the original WHAM! dataset, except noise, which was recorded spatialized. The anechoic sources (i.e., direct path signals) serve as targets to reverberated sources for models involving dereverberation, allowing them to be trained without needing to account for the time delay of the spatialized sources.
In spatializing the audio, we generated a two-channel version of the dataset, using microphone spacing from the WHAM! noise metadata, 
but in this study we focus on single-channel separation and use only the left channel.
The spatialized audio was rescaled to remove attenuation, such that the non-spatialized WHAM! and anechoic WHAMR! differ only by small time delays, and we found negligible performance differences when training and testing models using the two datasets.  While the results for non-reverberant conditions in Section~\ref{sec:results} use anechoic WHAMR!, they are directly comparable with WHAM!~\cite{wichern2019wham}.

Since all source, noise, and reverberated components and their combinations are included in the corpus, several enhancement, separation, and joint enhancement-separation tasks are enabled for training and evaluation. %
For example, in separating noisy and reverberant speech, we may want to produce either two clean, anechoic recordings or two clean, reverberant recordings, leaving dereverberation to post-processing.
We choose to define four core separation tasks:
\begin{itemize}
\setlength{\itemsep}{-1pt}
    \item \textbf{clean} -- anechoic clean mixture to anechoic sources
    \item \textbf{noisy} -- anechoic noisy mixture to anechoic sources
    \item \textbf{reverberant} -- reverberant clean mixture to anechoic sources
    \item \textbf{noisy and reverberant} -- reverberant noisy mixture to anechoic sources
\end{itemize}
All other configurations are only considered and evaluated as sub-components to the above tasks. Since each condition has its own unprocessed signal-to-distortion ratio~(SDR), comparisons across tasks can be difficult. By restricting to the above tasks, where the targets are the same in all four conditions, raw SDR can be thought of as a directly comparable, ``objective'' quality metric of the output sources across tasks. SDR \textit{improvement} also brings insight by reporting how much improvement a system has made to the signal. %

\section{Experimental Configurations}
\label{sec:conf}

\subsection{Network Configurations}
\label{ssec:net_config}

For our experiments, we use four basic network configurations, all under the same paradigm. %
First, the waveforms are projected to a spectro-temporal representation. Next, an internal network takes the spectral representation and produces a spectral mask with values from 0 to 1. %
Finally, this spectral mask is applied to the original representation, suppressing interfering signals, before the representation is projected back to produce an estimated source waveform. In enhancement, the internal masking network produces a single mask, attempting to suppress noise and/or reverberation. In separation, the masking network produces a mask for each speech signal, attempting to suppress the interfering speakers from each target speaker.

The four configurations we use are the possible combinations of two spectral feature extractors and two internal masking networks. The feature extractors we compare are a standard short-time Fourier transform~(STFT) and a TasNet-style learned basis transform~\cite{Luo2018,luo2019convTasNet}, which consists of projecting sliding-window subsegments of the waveform onto a set of learned basis functions. The resulting weights can be applied to a reconstruction set of basis functions and summed together along the same sliding window to reconstruct the signal under a similar paradigm to overlap-and-add for the STFT. For internal masking, we evaluate both bi-directional long short-term memory~(BLSTM) networks (the typical internals of earlier deep learning-based speech separation systems~\cite{Hershey2016,Isik2016,Kolbaek2017,Wang2018ICASSP04Alternative,Luo2018,wichern2019wham}) and temporal convolutional networks~(TCN)~\cite{Lea2017} with dilated convolutions (popular in recent state-of-the-art separation techniques~\cite{luo2019convTasNet,shi2019furcanext}).

For consistency with the prior WHAM! work \cite{wichern2019wham}, our BLSTM architecture has four BLSTM layers with 600 units in each direction followed by a fully-connected layer for each output mask. A dropout of 0.3 is applied on each BLSTM layer output except the last. The TCN architecture was chosen to match the best system reported in~\cite{luo2019convTasNet}. It consists of a 128-dimensional bottleneck, 128-dimensional skip-connection paths, and 512 channels in the convolutional blocks, with kernel size 3, 8 blocks per repeat, and 3 repeats.

The STFT features are also chosen to be consistent with \cite{wichern2019wham}, with a window length of 32~ms and hop size of 8~ms. The log of the magnitude spectrum is used as input to the internal masking network. The learned basis feature parameters are also chosen to be consistent with \cite{wichern2019wham}, with a 10~ms window and 5~ms hop, with 500 learned basis vectors.  While the original BLSTM TasNet~\cite{Luo2018} used a gated convolutional encoder, in this work we use a single learned encoder and ReLU nonlinearity as in Conv-TasNet~\cite{luo2019convTasNet} for both the BLSTM and TCN masking networks with learned bases.

For separation, we evaluate learned basis configurations only, as they have been shown to outperform STFT-based methods on clean data, and performed best in preliminary experiments.
However, %
we perform full comparisons of the differing features for enhancement, for which TasNet-like systems have only rarely been evaluated~\cite{luo2018dereverb}.

We train all networks using permutation invariant training~\cite{Hershey2016, Kolbaek2017} with the scale-invariant signal-to-distortion ratio (SI-SDR, also referred to as SI-SNR) waveform-level training objective~\cite{Isik2016,Luo2018,LeRoux2018SISDR}. SI-SDR is also the evaluation metric and allows for end-to-end joint training of cascaded enhancement and separation models:
\begin{gather}
\text{SI-SDR} = 10 \log_{10} ({\Vert \alpha s \Vert^2}/{\Vert \alpha s - \hat{s}\Vert^2}), \, \alpha = {\langle \hat{s},s \rangle}/{\Vert s \Vert^2}. \label{eq:sisnr}
\end{gather}
Because the loss is scale-invariant and the outputs are not constrained to sum up to the mixture, the outputs may be in a different dynamic range as the mixture, which as we will see can lead to problems with the cascaded models proposed in this work.

\subsection{Cascaded Models}
\label{ssec:cascade_models}

In addition to training single models for each of the WHAMR! core tasks, we evaluate combinations of models in which enhancement (i.e., denoising and/or dereverberation) and separation systems are cascaded, with the output of one system being fed into the next. %
The main motivation is that jointly separating and enhancing may be too difficult for a single network to learn, and modularization may allow the networks to focus on specific tasks.  Two-stage approaches have previously been explored for denoising plus dereverberation~\cite{han2015learning, zhao2018two}, separation plus dereverberation~\cite{delfarah2019deep}, and denoising plus separation~\cite{wichern2019wham}.

The cascaded configurations we consider consist of an optional pre-enhancement system cascaded into a separation network cascaded into an optional post-enhancement system. We evaluate all combinations where noise is removed by either pre-enhancement or the separator, and reverberation is removed by either pre-enhancement, post-enhancement, or the separator. Post-separation denoising is not considered, as separation-without-denoising is a somewhat ill-defined task: %
noise does not `belong' to either speech signal, so it is unclear how the network should distribute the noise when not removing it.

For cascaded systems, the sub-models are trained with appropriate input and targets for each  sub-task. %
For example, in the system consisting of denoising followed by separation then dereverberation, the networks are trained as follows: pre-enhancement is trained with noisy reverberant mixtures as input and clean reverberant mixtures as output; the separator with reverberant mixtures as input and reverberant sources as output; and post-enhancement with single reverberant sources as input and single anechoic sources as output.

As mentioned above, due to the scale-invariant loss function, each model's outputs have no constraint to be within any particular dynamic range, and we thus observe strong degradation in performance in cascaded systems when sub-models are trained separately, due to the scaling mismatch between the output of one model and the training data of the next. To address this problem, we scale each output $\hat{s}$, obtained from an input mixture $x$ as an estimate for a target source $s$, to make it consistent with the scaling of $s$ in $x$. Because $s$ is unknown, we need to rely on $\hat{s}$ and $x$ alone. If we assume that the interfering signal $n=x-s$ is orthogonal to $s$, which is generally approximately the case, and that the direction of $\hat{s}$ is close to that of $s$, then a reasonable choice for the rescaling factor $\beta(\hat{s}|x)$ is that obtained by ensuring that $\beta(\hat{s}|x)\hat{s}$ is orthogonal to the residual $\hat{n} = x - \beta(\hat{s}|x) \hat{s}$. %
This results in a scaling factor
\begin{gather}
\beta(\hat{s}|x) = \frac{\langle x,\hat{s}\rangle}{\Vert\hat{s}\Vert^2}.
\end{gather}
As the estimate $\hat{s}$ improves (i.e., $\hat{s}$ and $s$ become more colinear), the scaling factor improves as well.

When the best-performing system of a WHAMR! task is a cascaded model, we also evaluate the system with additional end-to-end tuning.
Since all component systems are waveform-to-waveform, we can tune the entire system by performing additional training through all cascaded sub-models directly.
End-to-end joint training of sub-models has been shown to be successful in joint training of automatic speech recognition with enhancement and separation~\cite{ochiai2017,Settle2018ICASSP04,seki2018purely,chang2019}.

\subsection{Training Configurations}
\label{ssec:train_config}

All networks are trained on 4 second segments using the Adam algorithm~\cite{Kingma2015Adam}. The learning rate is decreased by a factor of 2 if validation loss does not improve for 3 consecutive epochs. Gradient clipping is applied with a maximum $\ell_2$ norm of 5. Models are trained for 100 epochs with an initial learning rate of $10^{-3}$, with the exception of cascaded model tuning, during which we train the models for 25 epochs with a learning rate of $10^{-4}$. Because the SI-SDR loss is undefined for silent sources, training models on the \textit{max} data subset is cumbersome, as the 4 s segments randomly sampled during training occasionally fall within regions where only one speaker is talking. Thus, for the 16~kHz~\textit{max} condition, we train on 16~kHz~\textit{min}.  Unless otherwise noted, all results are for the 8~kHz~\textit{min} condition.

\vspace{-.2cm}
\section{Experimental Results}
\label{sec:results}
\vspace{-.1cm}

For all experiments, we report results using scale-invariant source-to-distortion ratio~(SI-SDR)~\cite{LeRoux2018SISDR}, which is also the training objective.  Furthermore, because the input SI-SDR between tasks is highly variable, we also report the SI-SDR improvement ($\Delta$), i.e., the difference between output and input SI-SDR.  %

\begin{table}[tbp]
\centering
\vspace{-.2cm}
\caption{SI-SDR [dB] results for a single separation network. Highlighted rows represent new WHAMR! conditions.}
\vspace{.1cm}
\label{table:base_separation}
\begin{adjustbox}{max width=0.9\columnwidth}
\sisetup{table-format=2.1,round-mode=places,round-precision=1,table-number-alignment = center,detect-weight=true,detect-inline-weight=math}
\begingroup
\renewcommand*{\arraystretch}{0.9}
\begin{tabular}{cc|S[table-format=2.1,table-number-alignment = center]|S[table-format=2.1,table-number-alignment = center]S|SS}
\toprule
\multicolumn{2}{c}{Input} & \multicolumn{1}{c}{} & \multicolumn{2}{c}{Conv-TasNet} & \multicolumn{2}{c}{TasNet-BLSTM} \\ \cmidrule(lr){1-2} \cmidrule(lr){4-5} \cmidrule(lr){6-7}
Noise & \multicolumn{1}{c}{Reverb} & \multicolumn{1}{c}{Input} & Output & \multicolumn{1}{c}{$\Delta$} & Output & {$\Delta$} \\ \midrule
 &  & 0.00 & 12.91 & 12.91 & \bfseries 14.16 & \bfseries 14.16 \\
{\checkmark} & {} & -4.49 & 7.01 & 11.50 & \bfseries 7.48 & \bfseries 11.97 \\ \rowcolor[HTML]{FDD49F}
 & {\checkmark} & -3.29 & 4.27 & 7.56 & \bfseries 5.58 & \bfseries 8.87 \\ \rowcolor[HTML]{FDD49F}
{\checkmark} & {\checkmark} & -6.13 & 2.22 & 8.34 & \bfseries 3.03 & \bfseries 9.16 \\ \bottomrule
\end{tabular}
\endgroup
 \end{adjustbox}
\vspace{-0.3cm}
\end{table}

\begin{table}[tbp]
\centering
\vspace{-.2cm}
\caption{\!$\Delta$SI-SDR [dB] comparison of our implementations with the best Conv-TasNet number in~\cite{luo2019convTasNet} and the corresponding learned feature configuration of 512~bases, window length~16, window shift~8.}
\vspace{.1cm}
\label{table:small_window}
\begin{adjustbox}{max width=0.75\columnwidth}
\sisetup{table-format=2.1,round-mode=places,round-precision=1,table-number-alignment = center,detect-weight=true,detect-inline-weight=math}
\begingroup
\renewcommand*{\arraystretch}{0.9}
\begin{tabular}{S[table-format=2.1,table-number-alignment = center]|S|S}
\toprule
\multicolumn{1}{c}{TasNet-BLSTM} & \multicolumn{1}{c}{Conv-TasNet} & \multicolumn{1}{c}{Conv-TasNet~\cite{luo2019convTasNet}} \\ \midrule
\bfseries 16.580 & 14.40 & 15.3 \\ \bottomrule
\end{tabular}
\endgroup \end{adjustbox}
\vspace{-.3cm}
\end{table}

Table~\ref{table:base_separation} shows the results of our core systems, without cascade. Reverberation seems to be more challenging than noise as reflected by the lower SI-SDR. While the noisy and clean conditions are comparable in terms of SI-SDR improvement, they still differ significantly in terms of raw SI-SDR. Interestingly, we observe consistently better performance from the BLSTM model over the TCN model, which is somewhat unexpected. Indeed, although the BLSTM contains many more parameters than the TCN, this result contradicts prior results in the literature \cite{Luo2018,luo2019convTasNet}. A comparison of clean separation models with a smaller basis window is shown in Table~\ref{table:small_window}, confirming that the performance difference is not due to the window parameters.

In addition, we note that the TasNet-BLSTM numbers in the first two rows are considerably better than the corresponding numbers in the original WHAM! paper \cite{wichern2019wham}. The newer network uses the same configuration, but is trained with more aggressive gradient clipping and stagnation learning rate adjustment, which supports the findings regarding training optimizer parameters reported in~\cite{luo2018dereverb, luo2019convTasNet}.

Table~\ref{table:denoise_both} shows experimental results with enhancement networks. We use denoising and dereverberation of two-speaker mixtures as a proxy for all other enhancement conditions. Since performance trends are consistent across these two tasks, we think this is reasonable evidence to conclude that the learned feature BLSTM model (TasNet-BLSTM) is the best architecture for enhancement.  While the learned basis TCN and BLSTM perform similarly, we see significant drops in performance moving from learned basis to STFT features. This suggests that the benefits shown in speech separation are also likely present in speech denoising and dereverberation.

\begin{table}[tbp]
\centering
\caption{SI-SDR [dB] for two-speaker enhancement tasks.}
\label{table:denoise_both}
\begin{adjustbox}{max width=0.9\columnwidth}
\sisetup{table-format=2.1,round-mode=places,round-precision=1,table-number-alignment = center,detect-weight=true,detect-inline-weight=math}
\begingroup
\renewcommand*{\arraystretch}{0.9}
\begin{tabular}{cc|S[table-format=2.1,table-number-alignment = center]S|SS}
\toprule
\multicolumn{2}{c}{Net} & \multicolumn{2}{c}{Denoise} & \multicolumn{2}{c}{Dereverb} \\ \cmidrule(lr){1-2} \cmidrule(lr){3-4} \cmidrule(lr){5-6}
\multicolumn{1}{c}{Feature} & \multicolumn{1}{c}{Processor} & Output & \multicolumn{1}{c}{$\Delta$} & Output & \multicolumn{1}{c}{$\Delta$} \\ \midrule
\multicolumn{1}{c}{Learned} & \multicolumn{1}{c|}{TCN} & 10.80 & 9.62 & 7.23 & 3.19 \\
\multicolumn{1}{c}{Learned} & \multicolumn{1}{c|}{BLSTM} & \bfseries 11.24 & \bfseries 10.06 & \bfseries 8.46 & \bfseries 4.42 \\
\multicolumn{1}{c}{STFT} & \multicolumn{1}{c|}{TCN} & 8.40 & 7.21 & 4.04 & 0.00 \\
\multicolumn{1}{c}{STFT} & \multicolumn{1}{c|}{BLSTM} & 9.54 & 8.36 & 5.89 & 1.84 \\ \midrule
\multicolumn{2}{c}{Input SI-SDR:} & \multicolumn{2}{c}{\num{1.19}} & \multicolumn{2}{c}{\num{4.03}} \\ \bottomrule
\end{tabular}
\endgroup
 \end{adjustbox}
\vspace{-0.3cm}
\end{table}

Table~\ref{table:combos} shows the results of the cascaded model experiments. In accordance with the previous results, all sub-models are TasNet-BLSTM models. We see that in general, moving the speech enhancement (i.e., denoising and/or dereverberation) tasks to a separate model from separation seems to help performance. From Tables~\ref{table:combos}(b) and (c), reverberation appears to be particularly difficult for the separation network to remove. We also see that removing reverberation post-separation is slightly better than pre-separation.  As two sources will not have the same room impulse response, the dual-source (pre-enhancement) dereverberation network would have to appropriately compensate for two reverberation patterns, while the single-source dereverberation (post-enhancement) network handles only one. The separator network likely has a harder time separating the still-reverberant speech, but this effect appears to be smaller than the difference in single- and double-source dereverberation.

\begin{table}[]
\centering
\caption{Comparison of cascaded models.  A dash indicates speech separation without denoising/dereverberation, while \batsu\, indicates no enhancement sub-model was used.  Results are sorted by increasing performance. The highlighted rows indicate the non-cascaded single-model baseline.}
\vspace{.1cm}
\label{table:combos}
\begin{adjustbox}{max width=0.74\columnwidth}
\subfloat[noisy condition]{%
\sisetup{table-format=2.1,round-mode=places,round-precision=1,table-number-alignment = center,detect-weight=true,detect-inline-weight=math}
\begingroup
\renewcommand*{\arraystretch}{0.9}
\begin{tabular}{ccS[table-format=2.1,table-number-alignment = center]S}
\toprule
\multicolumn{2}{c}{System} &  \multicolumn{2}{c}{\multirow{3}{*}{\begin{tabular}[c]{@{}ccc@{}} & SI-SDR & \\ \cmidrule{1-3}\end{tabular}}}  \\ \cmidrule(lr){1-2}
\multirow{2}{*}{\begin{tabular}[c]{@{}c@{}}Pre-Enh.\\ Removes\end{tabular}} & \multirow{2}{*}{\begin{tabular}[c]{@{}c@{}}Separate Speech\\ while Removing\end{tabular}}  & &  \\
 &  & Output & \multicolumn{1}{c}{$\Delta$} \\ \midrule \rowcolor[HTML]{FDD49F}
\multicolumn{1}{c|}{\batsu} & \multicolumn{1}{c|}{noise} & 7.48 & 11.97 \\
\multicolumn{1}{c|}{noise} & \multicolumn{1}{c|}{--} & \bfseries 8.10 & \bfseries 12.59 \\
\midrule
\multicolumn{2}{c}{Input SI-SDR:} & \multicolumn{2}{c}{\num{-4.49}} \\ \bottomrule
\end{tabular}
\endgroup
\label{subtable:combos_dn}}
\end{adjustbox}
\hfill
\begin{adjustbox}{max width=0.93\columnwidth}
\subfloat[reverberant condition]{%
\sisetup{table-format=2.1,round-mode=places,round-precision=1,table-number-alignment = center,detect-weight=true,detect-inline-weight=math}
\begingroup
\renewcommand*{\arraystretch}{0.9}
\begin{tabular}{cccS[table-format=2.1,table-number-alignment = center]S}
\toprule
\multicolumn{3}{c}{System} &  \multicolumn{2}{c}{\multirow{3}{*}{\begin{tabular}[c]{@{}ccc@{}} & SI-SDR & \\ \cmidrule{1-3}\end{tabular}}}  \\ \cmidrule(lr){1-3}
\multirow{2}{*}{\begin{tabular}[c]{@{}c@{}}Pre-Enh.\\ Removes\end{tabular}} & \multirow{2}{*}{\begin{tabular}[c]{@{}c@{}}Separate Speech\\ while Removing\end{tabular}} & \multirow{2}{*}{\begin{tabular}[c]{@{}c@{}}Post-Enh.\\ Removes\end{tabular}} & &  \\
 &  &  & Output & \multicolumn{1}{c}{$\Delta$} \\ \midrule \rowcolor[HTML]{FDD49F}
\multicolumn{1}{c|}{\batsu} & \multicolumn{1}{c|}{rev.} & \multicolumn{1}{c|}{\batsu} & 5.58 & 8.87 \\
\multicolumn{1}{c|}{rev.} & \multicolumn{1}{c|}{--} & \multicolumn{1}{c|}{\batsu} & 6.39 & 9.68 \\
\multicolumn{1}{c|}{\batsu} & \multicolumn{1}{c|}{--} & \multicolumn{1}{c|}{rev.} & \bfseries 6.59 & \bfseries 9.88 \\
\midrule
\multicolumn{1}{c}{\phantom{noise, re...}} &
\multicolumn{1}{c}{Input SI-SDR:} &
\multicolumn{1}{c}{} &
\multicolumn{2}{c}{\num{-3.29}} \\ \bottomrule
\end{tabular}
\endgroup
\label{subtable:combos_dr}}
\end{adjustbox}
\hfill
\begin{adjustbox}{max width=0.93\columnwidth}
\subfloat[noisy and reverberant condition]{%
\sisetup{table-format=2.1,round-mode=places,round-precision=1,table-number-alignment = center,detect-weight=true,detect-inline-weight=math}
\begingroup
\renewcommand*{\arraystretch}{0.9}
\begin{tabular}{cccS[table-format=2.1,table-number-alignment = center]S}
\toprule
\multicolumn{3}{c}{System} &  \multicolumn{2}{c}{\multirow{3}{*}{\begin{tabular}[c]{@{}ccc@{}} & SI-SDR & \\ \cmidrule{1-3}\end{tabular}}}  \\ \cmidrule(lr){1-3}
\multirow{2}{*}{\begin{tabular}[c]{@{}c@{}}Pre-Enh.\\ Removes\end{tabular}} & \multirow{2}{*}{\begin{tabular}[c]{@{}c@{}}Separate speech\\ while removing\end{tabular}} & \multirow{2}{*}{\begin{tabular}[c]{@{}c@{}}Post-Enh.\\ Removes\end{tabular}} & &  \\
 &  &  & Output & \multicolumn{1}{c}{$\Delta$} \\ \midrule \rowcolor[HTML]{FDD49F}
\multicolumn{1}{c|}{\batsu} & \multicolumn{1}{c|}{noise, rev.} & \multicolumn{1}{c|}{\batsu} & 3.03 & 9.16 \\
\multicolumn{1}{c|}{noise} & \multicolumn{1}{c|}{rev.} & \multicolumn{1}{c|}{\batsu} & 3.53 & 9.66 \\
\multicolumn{1}{c|}{noise, rev.} & \multicolumn{1}{c|}{--} & \multicolumn{1}{c|}{\batsu} & 3.56 & 9.69 \\
\multicolumn{1}{c|}{rev.} & \multicolumn{1}{c|}{noise} & \multicolumn{1}{c|}{\batsu} & 3.72 & 9.84 \\
\multicolumn{1}{c|}{\batsu} & \multicolumn{1}{c|}{noise} & \multicolumn{1}{c|}{rev.} & 3.66  & 9.78 \\
\multicolumn{1}{c|}{noise} & \multicolumn{1}{c|}{--} & \multicolumn{1}{c|}{rev.} & \bfseries 3.97 & \bfseries 10.10 \\
\midrule
\multicolumn{1}{c}{\phantom{noise, rev.}} &
\multicolumn{1}{c}{Input SI-SDR:} &
\multicolumn{1}{c}{} & 
\multicolumn{2}{c}{\num{-6.13}} \\ \bottomrule
\end{tabular}
\endgroup
\label{subtable:combos_dndr}}
\end{adjustbox}
\vspace{-0.8cm}
\end{table}

While the cascaded systems do have 2 or 3 times as many parameters as the non-cascaded system, this does not seem to be the sole source of performance improvement, as single models with increased numbers of BLSTM layers provided little performance gain over the results in Table~\ref{table:base_separation}.  Furthermore, training equivalent cascaded systems from scratch without individual pre-training of the pre-enhancement, separation, and post-enhancement stages %
provided noticeably less performance improvement over the single network results from Table~\ref{table:base_separation} than the reported cascaded systems in Table~\ref{table:combos}.

\begin{table}[tbp]
\centering
\caption{SI-SDR comparison of best models with and without additional training. Dashes indicate the best system was not cascaded.}
\vspace{.1cm}
\label{table:tuning}
\begin{adjustbox}{max width=0.95\columnwidth}
\sisetup{table-format=2.1,round-mode=places,round-precision=1,table-number-alignment = center,detect-weight=true,detect-inline-weight=math}
\begingroup
\renewcommand*{\arraystretch}{0.9}
\begin{tabular}{cc|S[table-format=2.1,table-number-alignment = center]|S[table-format=2.1,table-number-alignment = center]S|SS}
\toprule
 & \multicolumn{1}{c}{} & \multicolumn{1}{c}{} & \multicolumn{2}{c}{\multirow{2}{*}{\begin{tabular}[c]{@{}c@{}}Best System\\ w/o Tuning\end{tabular}}} & & \\
\multicolumn{2}{c}{Input} & \multicolumn{1}{c}{} & & \multicolumn{1}{c}{} & \multicolumn{2}{c}{Tuned} \\ \cmidrule(lr){1-2} \cmidrule(lr){4-5} \cmidrule(lr){6-7}
Noise & \multicolumn{1}{c}{Reverb} & \multicolumn{1}{c}{Input} & Output & \multicolumn{1}{c}{$\Delta$} & Output & \multicolumn{1}{c}{$\Delta$} \\ \midrule
& & 0.00 & 14.16 & 14.16 & {--} & {--} \\
{\checkmark} &  & -4.49 & 8.10 & 12.59 & 8.34 & 12.86 \\
 & {\checkmark} & -3.29 & 6.59 & 9.88 & 6.99 & 10.27 \\
{\checkmark} & {\checkmark} & -6.13 & 3.97 & 10.10 & 4.72 & 10.84 \\ \bottomrule
\end{tabular}
\endgroup
 \end{adjustbox}
\end{table}

Table~\ref{table:tuning} shows the results of tuning the cascaded systems with additional end-to-end training. Tuning the systems helps, although the performance gains are minor. The noisy and reverberant system, which contains three sub-models in contrast to the others with two, shows the greatest improvement. This suggests training helps with improving the coupling of the connected models.

\begin{table}[tbp]
\centering
\caption{SI-SDR evaluation of 16 kHz conditions using the best model configuration trained on the 16 kHz \textit{min} subset.}
\vspace{.1cm}
\label{table:16k}
\begin{adjustbox}{max width=\columnwidth}
\sisetup{table-format=2.1,round-mode=places,round-precision=1,table-number-alignment = center,detect-weight=true,detect-inline-weight=math}
\begingroup
\renewcommand*{\arraystretch}{0.9}
\begin{tabular}{cc|S[table-format=2.1,table-number-alignment = center]|S[table-format=2.1,table-number-alignment = center]S|S|SS}
\toprule
\multicolumn{2}{c}{Input} & \multicolumn{3}{c}{16 kHz Min} & \multicolumn{3}{c}{16 kHz Max} \\ \cmidrule(lr){1-2}\cmidrule(lr){3-5} \cmidrule(lr){6-8} 
Noise & \multicolumn{1}{c}{Reverb} & \multicolumn{1}{c}{Input} & Output & \multicolumn{1}{c}{$\Delta$} & \multicolumn{1}{c}{Input} & Output & \multicolumn{1}{c}{$\Delta$} \\ \midrule
 &  & 0.00 & 12.86 & 12.86 & 0.00 & 12.71 & 12.71 \\
{\checkmark} &  & -4.57 & 7.79 & 12.36 & -5.84 & 7.47 & 13.32 \\
 & {\checkmark} & -3.30 & 5.63 & 8.93 & -3.41 & 5.39 & 8.81 \\
{\checkmark} & {\checkmark} & -6.19 & 3.74 & 9.94 & -7.20 & 3.50 & 10.70 \\ \bottomrule
\end{tabular}
\endgroup
 \end{adjustbox}
 \vspace{-0.1cm}
\end{table}

Table~\ref{table:16k} shows the results of our 16 kHz systems. As mentioned earlier, we trained on 16 kHz \textit{min} and evaluated on both the \textit{min} and \textit{max} conditions. Although the performance on 16 kHz data is worse than in the 8 kHz systems, there does not appear to be any significant breakdown in performance. Similarly, performance in the \textit{max} condition is only slightly worse than the \textit{min} condition. %
Although the SI-SDR improvement in the noisy case is better in \textit{max} than \textit{min}, this is likely due to differences in amount of speech and does not reflect any significant difference in performance.

\section{Conclusion}
\label{sec:conclusion}

We have introduced WHAMR!, an extension of the WHAM! noisy speech separation dataset to include reverberation, with the goal of further promoting the advancement of speech separation technologies towards more realistic conditions. Preliminary results demonstrate that, although noise and reverberation do degrade overall performance, networks with learned basis feature representations are effective not only in separation but also in speech enhancement. We have also demonstrated the value in using cascaded models combining pre-trained separation and enhancement modules, and of further jointly fine-tuning them, establishing strong baseline results for the WHAMR! dataset.  Extending the proposed model cascades to stereo is an important topic of future work, and is supported in the WHAMR! scripts available at \url{http://wham.whisper.ai}. 

\vfill\pagebreak

\bibliographystyle{IEEEtran}
\bibliography{refs}

\end{document}